\documentclass[a4paper]{jpconf}

\usepackage{graphicx}

\usepackage{epsfig}

\usepackage[ngerman, english]{babel}
\usepackage{cernlatex/cernunits,cernlatex/cernchemsym,cernlatex/heppennames2}
\usepackage{mcite}

\RequirePackage{include/iopams}

\providecommand*\snn{\sqrt{s_{\rm NN}}\xspace}

\providecommand*\gevc{GeV/$c$\xspace}

\providecommand{\s}{$\sqrt{s}$\xspace}

\providecommand{\dedx}{d$E$/d$x$\xspace}

\graphicspath{{.\fig/}}

\begin{document}
\title{Measurement of electrons from heavy-flavour hadron decays in pp and p-Pb collisions with ALICE}

\author{M Heide (for the ALICE collaboration)}

\address{Institut f\"ur Kernphysik, Wilhelm-Klemm-Str. 9, 48149 M\"unster, Germany}

\ead{mheide@wwu.de}

\begin{abstract}
ALICE measured electrons from inclusive heavy-flavour hadron decays and beauty-hadron decays at mid-rapidity in pp collisions at \s = 7 TeV and 2.76 TeV and p-Pb collisions at $\snn$ = 5.02 TeV. For both pp collision energies, $p_{\rm T}$-differential electron production cross sections are presented and compared to pQCD predictions. 
For p-Pb collisions, the $p_{\rm T}$-dependent nuclear modification factor of electrons from inclusive heavy-flavour hadron decays is presented.
\\
\\
(Some figures in this article are in colour only in the electronic version)
\end{abstract}

\section{Introduction}
\label{sec:intro}
Measurements of 
electrons from heavy-flavour hadron decays in pp collisions test 
perturbative QCD (pQCD) predictions and are 
references for \mbox{Pb-Pb} collisions, where heavy quarks are probes for in-medium energy loss 
\cite{eloss}. 
Cold nuclear matter effects 
can be disentangled from hot-medium effects via 
measurements of heavy-flavour hadron decay electrons in p-Pb collisions \cite{Salgado}.

The large branching ratio of semi-electronic charm and beauty-hadron decays ($\approx$ 10\%) makes electron measurements an alternative to analyses of heavy-flavour production via reconstruction of exclusive hadronic decays. With the ALICE \cite{alice} experiment inner barrel detectors, electrons can be identified 
down to low transverse momenta $p_{\rm T}$. Moreover, electrons from beauty-hadron decays can be measured separately, exploiting the spatial resolution provided by the ALICE Inner Tracking System (ITS) to select single electrons or secondary vertices with large displacements from the primary vertex. 
In addition, the ratio of electrons from beauty to those from inclusive heavy-flavour decays can be determined via azimuthal angular electron-hadron correlations.


\section{Electron identification}
\label{sec:eID}
All analyses of electrons from heavy-flavour hadron decays select a pure electron sample at mid-rapidity, using signals from the Time Projection Chamber (TPC) and from at least one of the devices Time of Flight (TOF), Electromagnetic Calorimeter (EMCal), and Transition Radiation Detector (TRD), with a track quality selection using ITS and TPC tracking information.
TPC particle identification is based on energy deposition per unit of length, \dedx, while TOF relates particle mass hypotheses to the time of flight and momentum $p$. The EMCal identifies electrons by their ratio $E/p \approx 1$, with the energy $E$ deposited by an absorbed particle. Electron-hadron separation in the TRD uses 
a signal composed of the energy deposited by a charged particle in the TRD gas volume and the transition radiation emitted by electrons with \mbox{$p \gtrsim$ 1 \gevc}.



\section{Additional requirements of beauty-hadron decay analyses}
\label{sec:beauty}
Kinematical differences between decays of $B$ and $D$ mesons can be used to identify electrons from beauty-hadron decays. Owing to the large decay length of $B$ mesons \mbox{($c\tau \approx$ 500 ${\rm \mu}$m)}, their decay electron tracks and vertices are on average further displaced from the primary vertex than others. The 
ALICE ITS is well suited to detect this difference. 
Moreover, the reconstruction of simulated events  shows that the near-side peak of azimuthal angular correlations between electrons from beauty decays and charged hadrons is about twice as wide as for charm decays.
In ALICE, electrons from beauty-hadron decays were measured via three approaches: 
\begin{itemize}
\item selection of electrons with a large transverse impact parameter to the primary vertex \cite{b7}. 
\item reconstruction of electron-hadron vertices displaced from the primary vertex, with requirements on the invariant mass of the electron-hadron pair 
and on the hadron $p_{\rm T}$.
\item measurement of azimuthal angular correlations between electrons and charged hadrons. The near-side peak in the angular distribution around the electron track is fitted with PYTHIA templates for beauty and charm decays to determine their relative contribution.
\end{itemize}

\section{Electron background subtraction and corrections}
\label{sec:eBg}

All selected electron samples 
contain electron background from non-heavy-flavour decays, which is subtracted by two alternative methods.  
In the first, $p_{\rm T}$-differential non-heavy-flavour 
electron spectra are calculated based on hadron cross sections measured at the LHC. 
For the analyses of beauty decays via impact parameter and displaced secondary vertices, simulations determine the fraction of electron background remaining after all selection criteria. All measured 
spectra are corrected for detector acceptances and selection efficiencies and undergo an unfolding procedure to correct for bremsstrahlung effects and momentum smearing in the reconstruction.

The second background subtraction method is based on the identification of photonic electrons by reconstruction of $e^+-e^-$ pairs with small invariant mass. The combinatorial background to this measurement is estimated via reconstruction of like-sign electron pairs and subtracted from the photonic electron yield. The resulting background is corrected for the pair-finding efficiency. This method is used in analyses of the relative beauty contributions to inclusive electron spectra from heavy-flavour hadron decays via electron-hadron correlations.



\section{Results from proton-proton collisions at \s = 7 TeV and \s = 2.76 TeV}
\label{sec:pp}
\begin{figure}[t]
\begin{minipage}[left]{0.32\textwidth}
\centering
\includegraphics[width=1.0\linewidth]{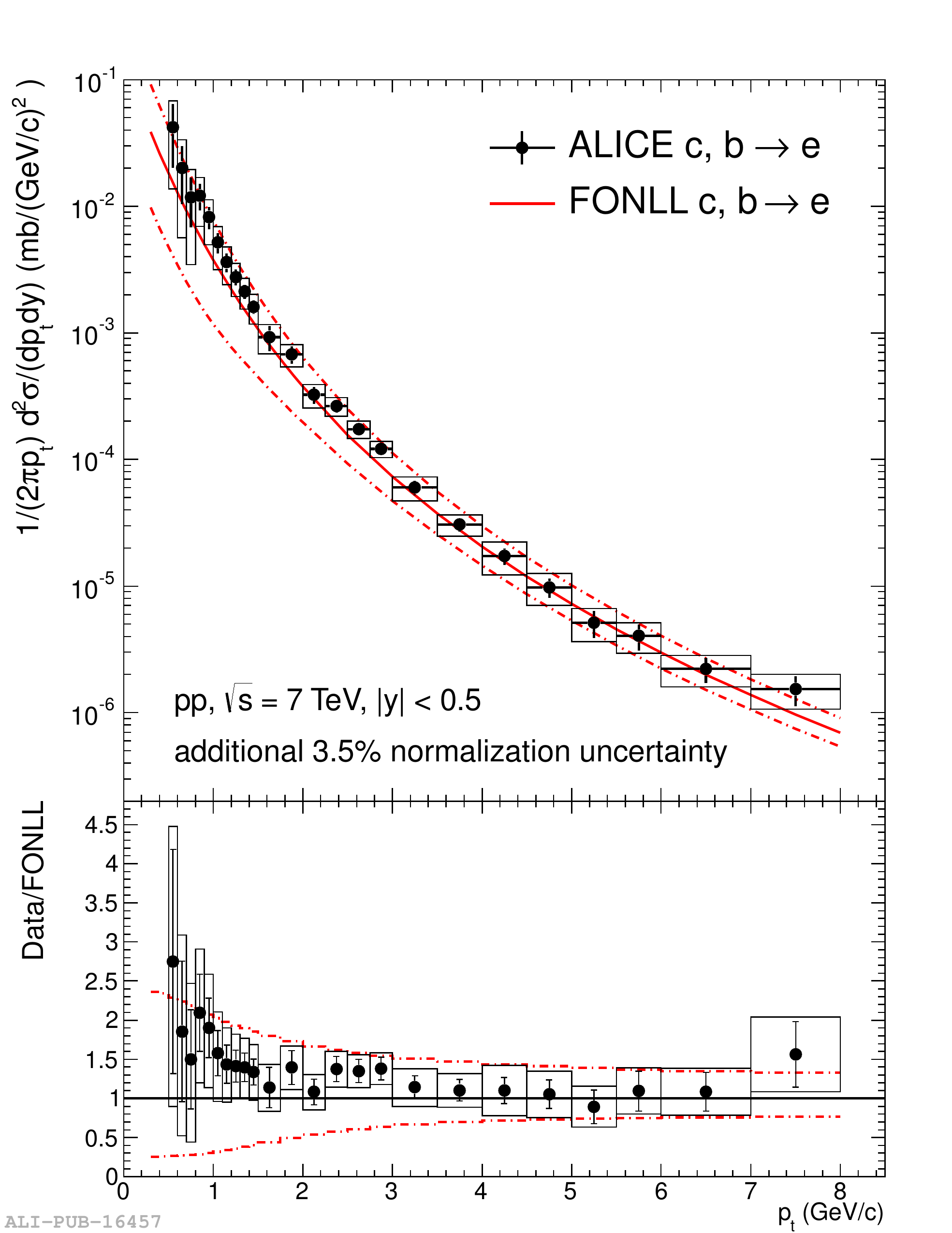}
\end{minipage}
\begin{minipage}[center]{0.34\textwidth}
\centering
\includegraphics[width=1.0\linewidth]{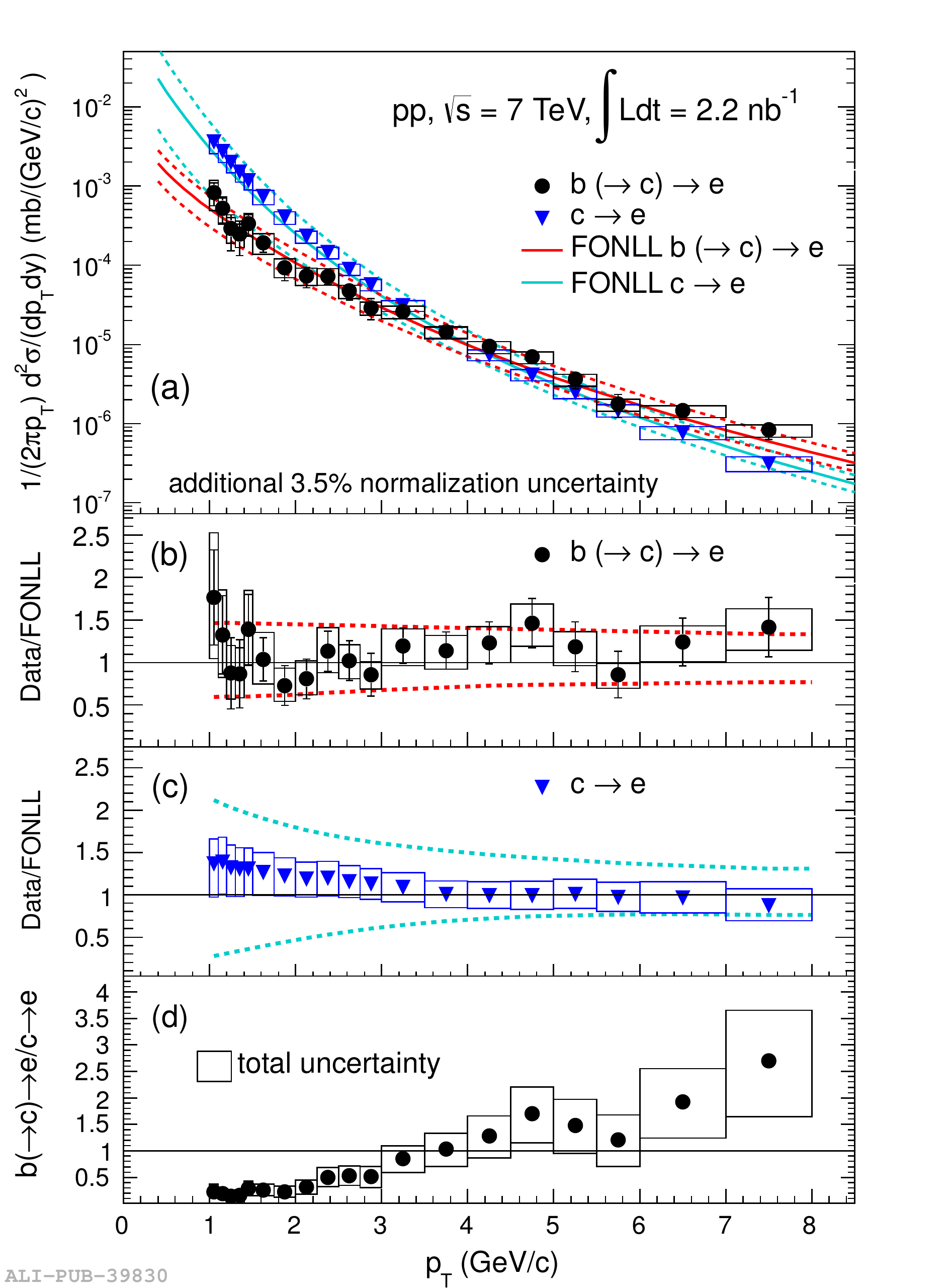}
\end{minipage}
\begin{minipage}[right]{0.34\textwidth}
\centering
\includegraphics[width=1.0\linewidth]{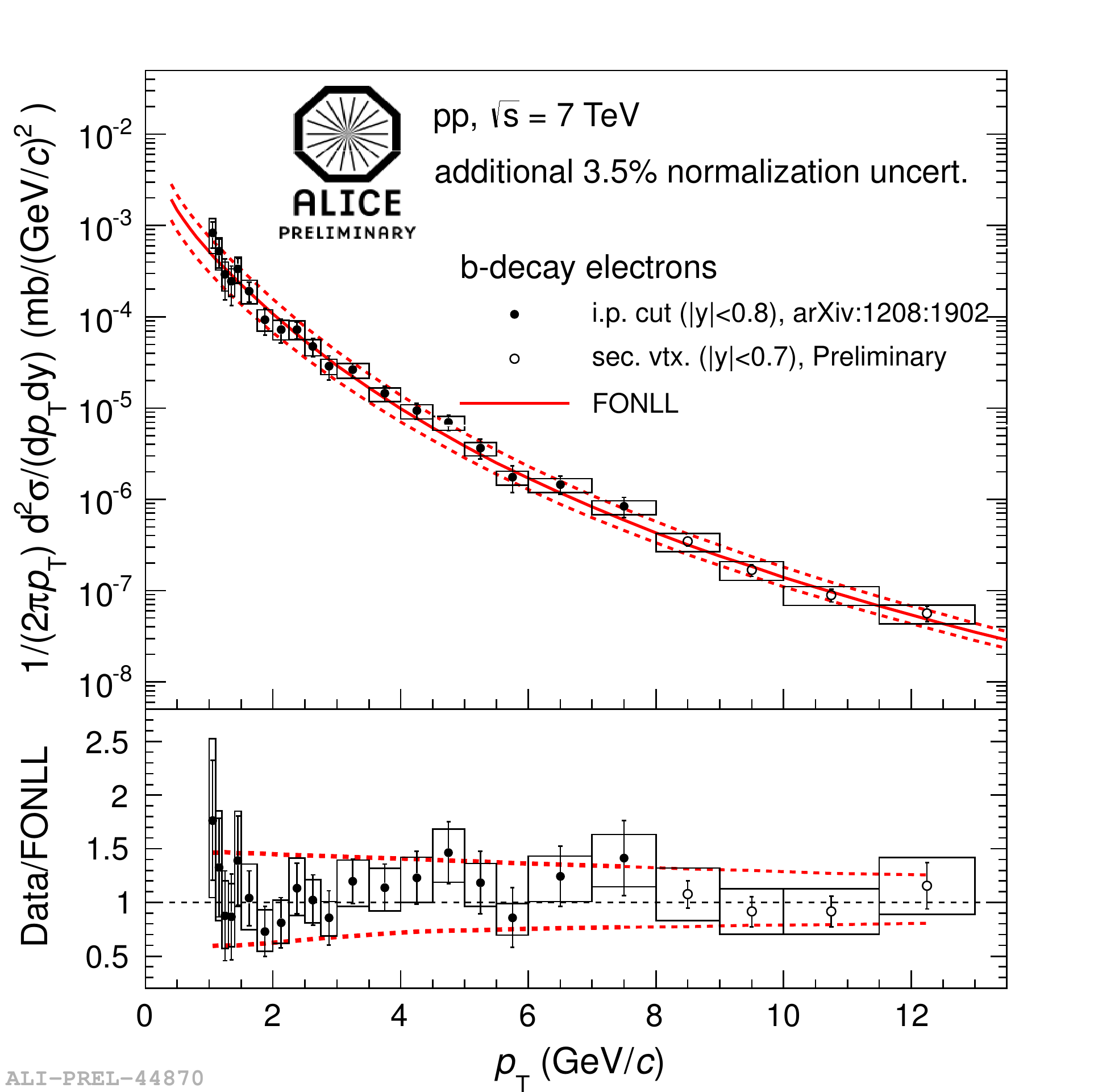}
\includegraphics[width=0.92\linewidth]{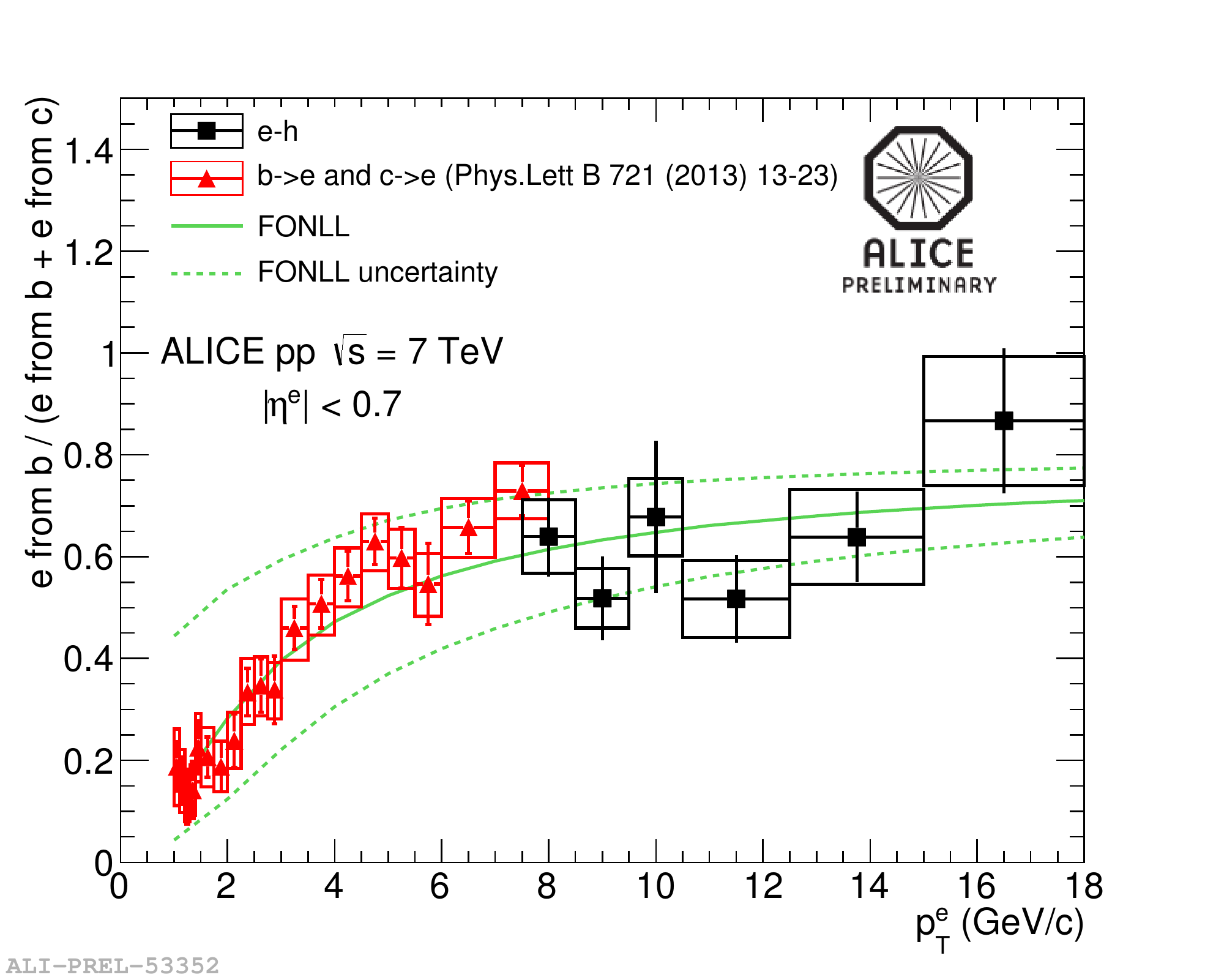}
\end{minipage}
\caption{Electron cross section from inclusive heavy-flavour hadron decays \cite{inc7} (left), beauty-hadron decays measured via impact parameter \cite{b7}, and charm-hadron decays (middle) in pp collisions at \mbox{\s = 7 TeV}. Upper right: electron cross section from beauty-hadron decays from \cite{b7}, with preliminary result of a measurement via displaced secondary vertices for \mbox{$p_{\rm T} > 8$ \gevc}. Lower right: ratio of electron yield from beauty-hadron decays to inclusive heavy-flavour hadron decays, measured via azimuthal angular electron-hadron correlations (black) and as ratio of results from middle and left panel (red). All results are compared to FONLL calculations \cite{fonll, *fonllp}.}
\label{fig:7tev}
\end{figure}

Fig. \ref{fig:7tev} shows measurements of heavy-flavour decay electrons in pp collisions at \mbox{\s = 7 TeV}. The left panel presents the $p_{\rm T}$-differential electron cross section from heavy-flavour hadron decays for 0.5 \gevc $< p_{\rm T} < 8$ \gevc in the rapidity range $|y| < 0.5$ \cite{inc7}. 
In the middle and right panel, results from all three analyses of beauty decay electrons are summarized: the $p_{\rm T}$-differential cross section determined by the impact parameter method is shown for \mbox{1 \gevc $< p_{\rm T} < 8$ \gevc} and $|y| < 0.8$ \cite{b7}. In comparison to the electron cross section from charm decays, calculated from ALICE measurements of charm-hadron cross sections, beauty decays become dominant for \mbox{$p_{\rm T} \gtrsim$ 4 \gevc}. The measurements of the $p_{\rm T}$-differential cross section via displaced electron-hadron vertices and of the ratio $e(b)/e(b+c)$ of electrons from beauty to inclusive heavy-flavour decays via electron-hadron correlations, both for $|y| < 0.7$, are complementary to the first method. Their results are comparable and extend the $p_{\rm T}$ range of beauty decay measurements to 13 and 18 \gevc, respectively. All spectra are described by pQCD predictions at fixed-order plus next-to-leading logarithms (FONLL \cite{fonll, *fonllp}) within uncertainties. Predictions from the $k_{\rm T}$-factorization \cite{kt, *ktp} and GM-VFN scheme \cite{gmv, *gmvp} achieve similarly good descriptions of the data.

\begin{figure}[t]
\begin{minipage}[left]{0.26\textwidth}
\centering
\includegraphics[width=1.0\linewidth]{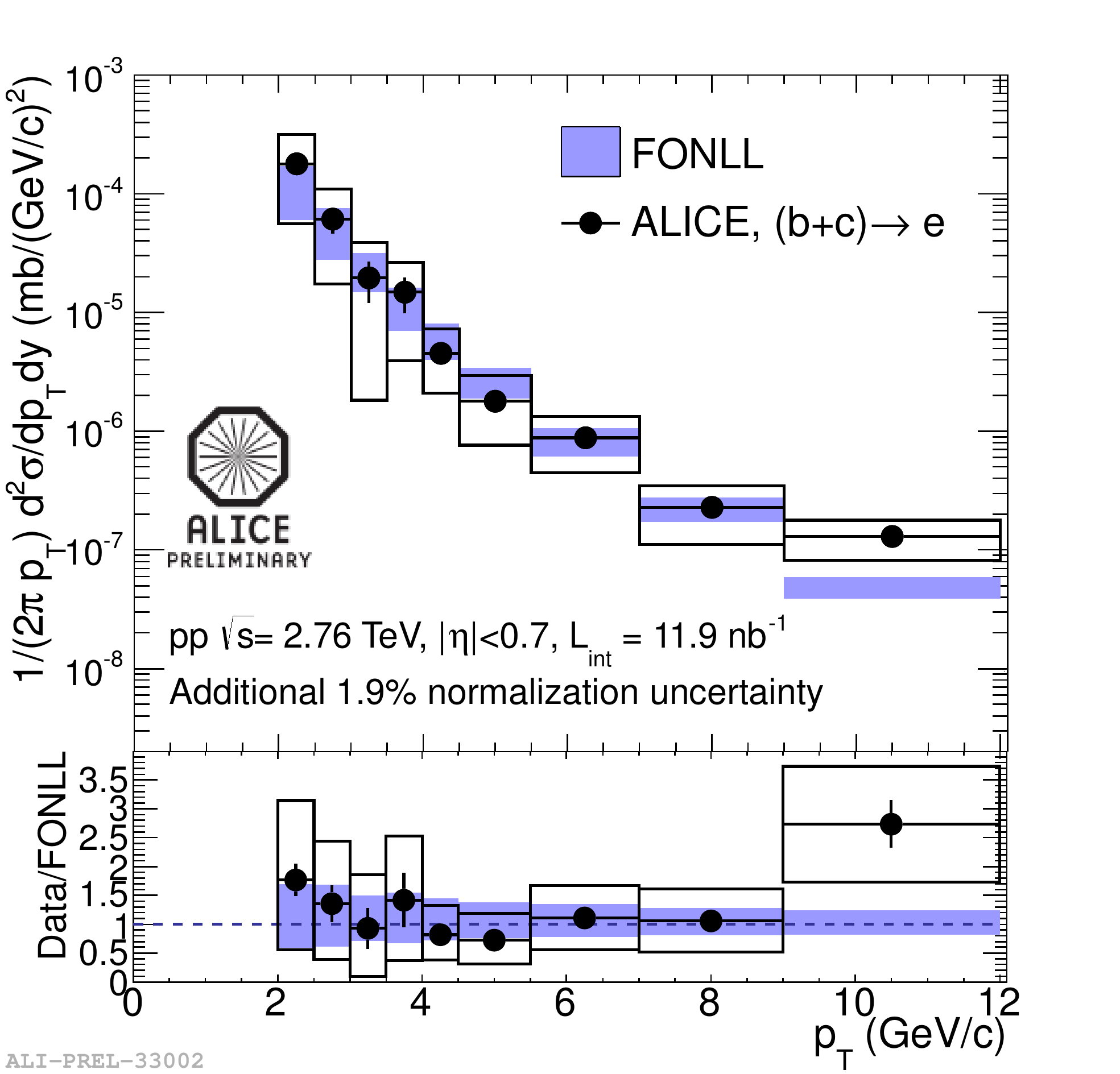}
\end{minipage}
\begin{minipage}[center]{0.34\textwidth}
\centering
\includegraphics[width=1.0\linewidth]{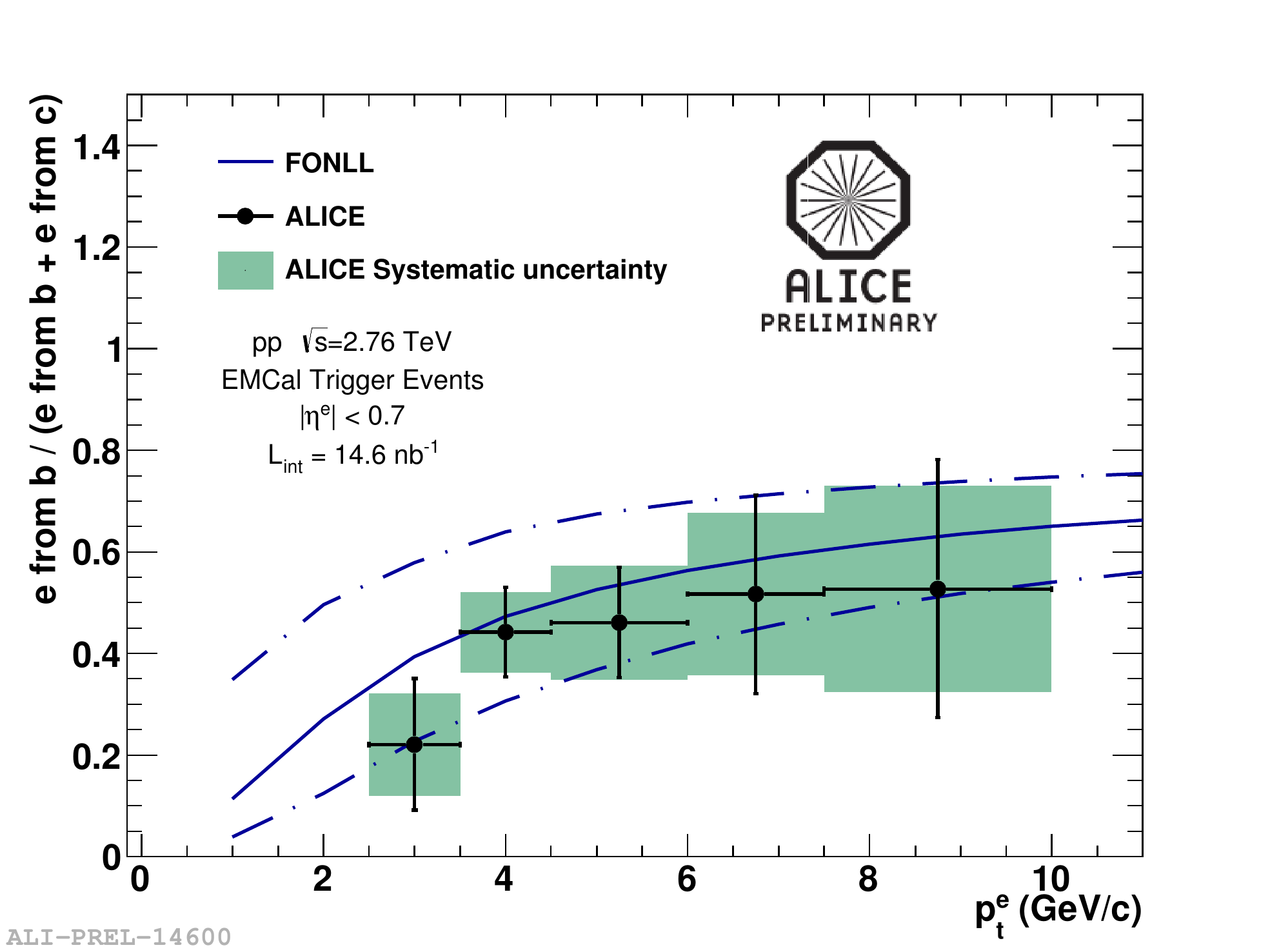}
\end{minipage}
\begin{minipage}[right]{0.40\textwidth}
\caption{Electron cross section from heavy-flavour hadron decays (left) and ratio of electrons from beauty to inclusive heavy-flavour hadron decays (right) in pp collisions at \mbox{\s = 2.76 TeV}, compared to FONLL predictions \cite{fonll, *fonllp}.}
\label{fig:pp276tev}
\end{minipage}
\end{figure}

At \s = 2.76 TeV, the reference energy for 
Pb-Pb collisions, the $p_{\rm T}$-differential 
electron cross section from inclusive heavy-flavour hadron decays was measured for \mbox{2 \gevc $< p_{\rm T} < 12$ \gevc} and $|y| < 0.7$ 
(Fig. \ref{fig:pp276tev}, left). For the same rapidity range, the ratio $e(b)/e(b+c)$ was determined via azimuthal angular electron-hadron correlations 
for \mbox{2.5 \gevc $< p_{\rm T} <$ 10 \gevc} (Fig. \ref{fig:pp276tev}, right). All measurements are described by FONLL predictions \cite{fonll, *fonllp}, calculations from the $k_T$-factorization framework \cite{kt, *ktp}, and the GM-VFN scheme \cite{gmv, *gmvp} within uncertainties.

\section{Results from p-Pb collisions at $\snn$ = 5.02 TeV}
\label{sec:pPb}

Inclusive heavy-flavour decay electrons were measured in p-Pb collisions for \mbox{$-0.14 < y < 1.06$} via two electron identification methods: in the $p_{\rm T}$ range from 0.5 \gevc to \mbox{6 \gevc}, TPC and TOF were utilized, while TPC and EMCal were used between 2 \gevc and 12 \gevc.  
The $p_{\rm T}$-differential cross sections measured by both methods agree in the common range, where the TPC-TOF result has smaller uncertainties and is therefore considered for the preliminary result. 

\begin{figure}[t]
\begin{minipage}[left]{0.5\textwidth}
\centering
\includegraphics[width=0.99\linewidth]{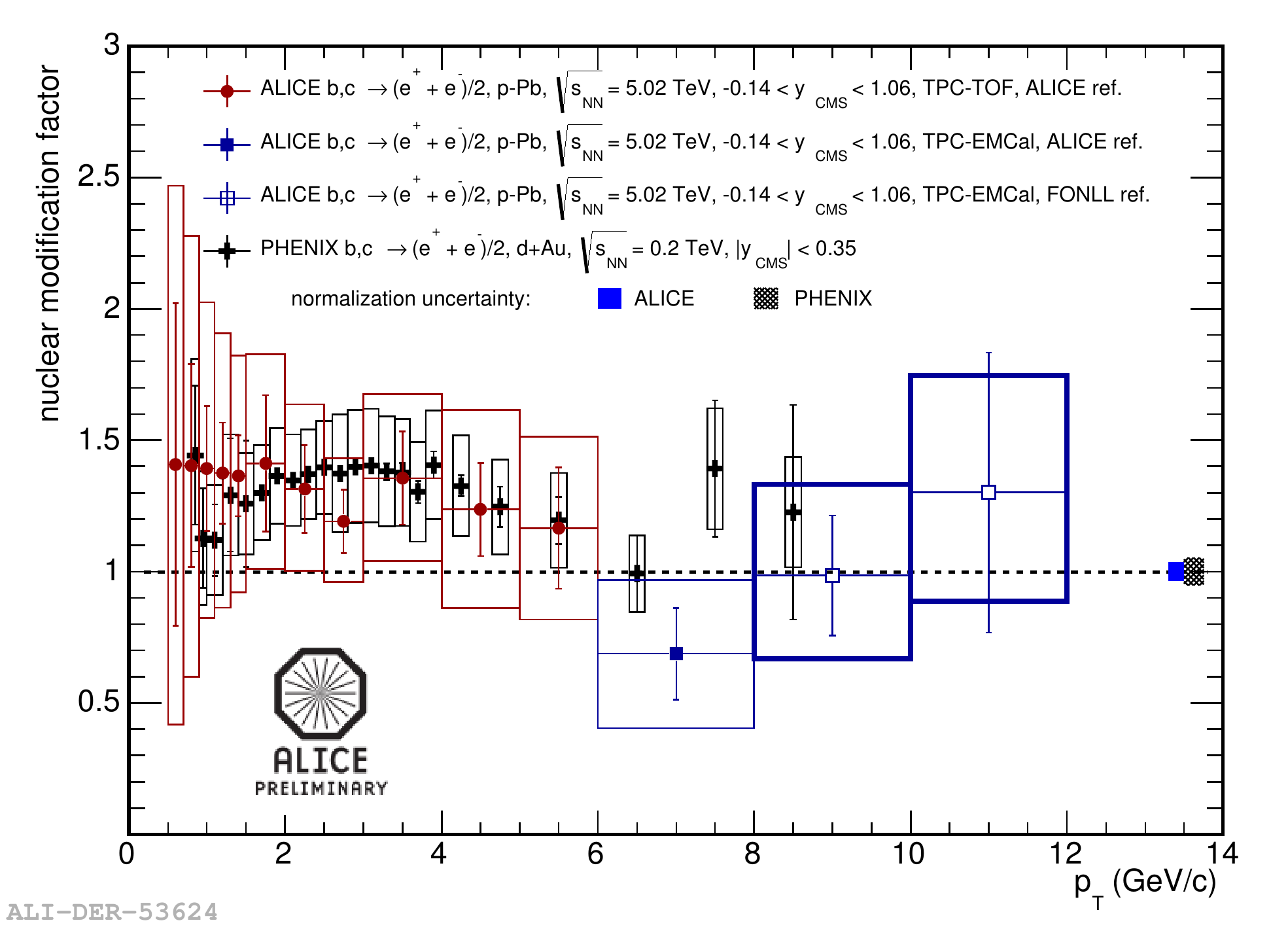}
\end{minipage}
\begin{minipage}[right]{0.5\textwidth}
\centering
\includegraphics[width=0.99\linewidth]{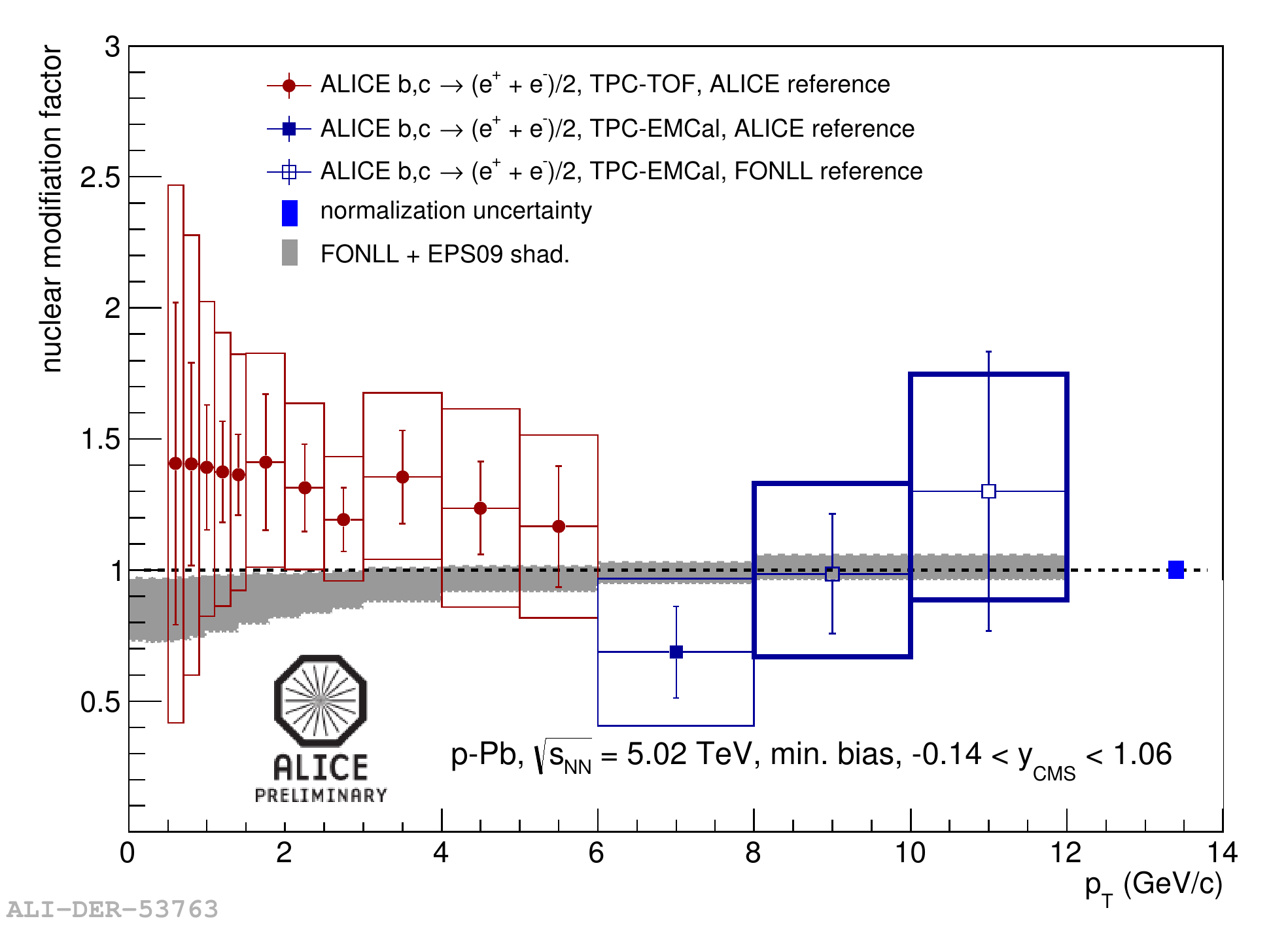}
\end{minipage}
\caption{$R_{\rm pPb}$ of electrons from inclusive heavy-flavour decays, compared to PHENIX results from \mbox{d-Au} collisions at \mbox{0.2 TeV} \cite{phenix} (left) and EPS09 nuclear shadowing predictions \cite{eps} (right).}
\label{fig:pPb}
\end{figure}

Fig. \ref{fig:pPb} shows the nuclear modification factor $R_{\rm pPb} (p_{\rm T}) = \frac{1}{\langle T_{\rm pPb}\rangle} \frac{{\rm d}N_{\rm pPb}/{\rm d}p_{\rm T}}{{\rm d}\sigma_{\rm pp}/{\rm d}p_{\rm T}}$ for electrons from heavy-flavour hadron decays. $\langle T_{\rm pPb} \rangle$ is the average nuclear overlap function, ${\rm d}N_{\rm pPb}/{\rm d}p_{\rm T}$ the $p_{\rm T}$-differential electron yield in p-Pb collisions, and ${\rm d}\sigma_{\rm pp}/{\rm d}p_{\rm T}$ the $p_{\rm T}$-differential cross section in pp collisions.
The electron spectrum in p-Pb collisions is compared to a pp reference spectrum that for $p_{\rm T} < 8$ \gevc  is based on the result from pp collisions at \mbox{\s = 7 TeV} (see \mbox{Fig. \ref{fig:7tev})}, scaled to \mbox{\s = \mbox{5.02 TeV}} using FONLL calculations. At higher $p_{\rm T}$, 
FONLL predictions were used as a reference. The $p_{\rm T}$ dependence of $R_{\rm pPb}$ is similar to the behaviour observed for non-photonic electrons at RHIC in d-Au collisions at \mbox{$\snn$ = 0.2 TeV} \cite{phenix} (Fig. \ref{fig:pPb}, left), which showed an enhancement at intermediate $p_{\rm T}$. However, the uncertainties of the ALICE measurements also allow for an agreement with predictions for the nuclear modification of the parton distribution functions from EPS09 parameterizations \cite{eps} that do not imply such an enhancement (Fig. \ref{fig:pPb}, right). Future reductions of these uncertainties are expected from an improved pp reference, interpolating with measurements at \s = 2.76 TeV, and the usage of a TRD electron trigger.

\section{Summary}
ALICE measured the $p_{\rm T}$-differential production cross sections of electrons from inclusive heavy-flavour and beauty-hadron decays at mid-rapidity for pp collisions at \mbox{\s = 7 TeV} and \mbox{\s = 2.76 TeV}. Results from three independent analyses of beauty-hadron decays are in mutual agreement. All measured cross sections from pp collisions are described by pQCD calculations within uncertainties. In p-Pb collisions at $\snn$ = 5.02 TeV, the $p_{\rm T}$ dependence of the nuclear modification factor of electrons from inclusive heavy-flavour hadron decays was determined. Within uncertainties, it agrees with the results from d-Au collisions at $\snn$ = 0.2 TeV by the PHENIX collaboration and with EPS09 nuclear shadowing predictions.

\section*{References}

\bibliographystyle{iopart-num}
\bibliography{HFEpp_pPb}

\end{document}